\documentclass[twocolumn,superscriptaddress,aps]{revtex4}
\usepackage{amsmath}
\usepackage{amssymb}
\usepackage{graphicx}
\usepackage{color}
\usepackage[colorlinks=true,linkcolor=blue,citecolor=blue]{hyperref}

\begin{document}

\title{Competing magnetic interactions in the antiferromagnetic topological insulator MnBi$_{2}$Te$_{4}$}

\author{J.-Q.~Yan}
\affiliation{Oak Ridge National Laboratory, Oak Ridge, TN, 37831, USA}

\author{D.~Pajerowski}
\affiliation{Oak Ridge National Laboratory, Oak Ridge, TN, 37831, USA}

\author{Liqin Ke}
\affiliation{Ames Laboratory, Ames, IA, 50011, USA}

\author{A.-M.~Nedi\'c}
\affiliation{Department of Physics and Astronomy, Iowa State University, Ames, IA, 50011, USA}

\author{Y.~Sizyuk}
\affiliation{Ames Laboratory, Ames, IA, 50011, USA}

\author{Elijah Gordon}
\affiliation{Ames Laboratory, Ames, IA, 50011, USA}

\author{P.~P.~Orth}
\affiliation{Ames Laboratory, Ames, IA, 50011, USA}
\affiliation{Department of Physics and Astronomy, Iowa State University, Ames, IA, 50011, USA}

\author{D.~Vaknin}
\affiliation{Ames Laboratory, Ames, IA, 50011, USA}
\affiliation{Department of Physics and Astronomy, Iowa State University, Ames, IA, 50011, USA}

\author{R.~J.~McQueeney}
\affiliation{Ames Laboratory, Ames, IA, 50011, USA}
\affiliation{Department of Physics and Astronomy, Iowa State University, Ames, IA, 50011, USA}

 \date{\today}

\begin{abstract}
The antiferromagnetic (AF) compound MnBi$_{2}$Te$_{4}$ is suggested to be the first realization of an antiferromagnetic (AF) topological insulator.  Here we report on inelastic neutron scattering studies of the magnetic interactions in MnBi$_{2}$Te$_{4}$ that possess ferromagnetic (FM) triangular layers with AF interlayer coupling.  The spin waves display a large spin gap and pairwise exchange interactions within the triangular layer are frustrated due to large next-nearest neighbor AF exchange.  The degree of frustration suggests proximity to a variety of magnetic phases, potentially including skyrmion phases, that could be accessed in chemically tuned compounds or upon the application of symmetry-breaking fields.
 \end{abstract}

\maketitle

The breaking of time-reversal symmetry by the introduction of magnetism in topological materials is key to unlocking unique topologically protected transport phenomena \cite{Tokura19}. For example, the quantum anomalous Hall effect has been demonstrated at low temperatures by inducing bulk ferromagnetism (FM) through the substitution of dilute magnetic ions, such as Cr or V, into (Bi,Sb)$_2$(Se,Te)$_{3}$ topological insulators \cite{Zhang13, Chang13, Chang15}.  While this is an incredibly important discovery, the disorder and inhomogeneity associated with these dilute FM systems present an obstacle to delivering quantum topological transport at routinely accessible temperatures. An alternate route to access these phenomena is to develop a new class of stoichiometric magnetic topological materials. MnBi$_{2}$Te$_{4}$ may be the first example of a stoichiometric antiferromagnetic topological insulator (AFTI) \cite{Otrokov18,Zhang18,Otrokov18_2,Lee18,Otrokov19,Yan19}.  AFTI are predicted to provide a platform for novel topological phases, such as quantum anomalous Hall insulators, axion insulators, or Weyl semimetals \cite{Moore10}. The symmetry, strength, and anisotropy of the magnetic interactions in AFTIs are important factors that control access to these quantum topological states.

MnBi$_{2}$Te$_{4}$ is a closely related structural variant of the tetradymite topological insulators, such as  Bi$_{2}$Te$_{3}$.  Whereas the teteradymite structure consists of stacked Te-Bi-Te-Bi-Te triangular (quintuple) layers, MnBi$_{2}$Te$_{4}$ consists of Te-Bi-Te-Mn-Te-Bi-Te septuple layers.  The electronic topology of inverted Bi--Te bands found in the tetradymites remains intact in MnBi$_{2}$Te$_{4}$ while the Mn triangular layers host large $S=$ 5/2 magnetic moments.  The AF ordering of Mn moments consists of FM triangular layers with AF interlayer coupling, referred to as $A$-type AF order, with moments pointing perpendicular to the layers \cite{Yan19}. The $A$-type structure provides access to novel topological phases via thin film growth with odd (time-reversal symmetry breaking) or even ($Z_{2}$ invariant) septuple layers \cite{Gong18,Otrokov19}.  In addition, relatively weak-field metamagnetic transitions allow access to canted, spin-flopped, or fully polarized magnetic structures \cite{Otrokov18_2,Lee18,Yan19}.  This flexibility of the magnetic structure has been utilized to demonstrate the QAH effect \cite{Zhang19}.

In this Letter, inelastic neutron scattering (INS) measurements on MnBi$_{2}$Te$_{4}$ reveal its Ising-like nature and relatively strong interlayer exchange interactions with large lifetime broadening.  We find that the next-nearest neighbor AF interaction ($J_{2}$) competes with nearest-neighbor FM interaction ($J_{1}$) within the triangular layer, placing the system close to the classical stability limit for intralayer FM correlations $|J_{2}/J_{1}| < 1/3$ \cite{Tanaka76,Murao96}.  These experimental observations are supported by first-principles (DFT+$U$) calculations of the magnetic interactions which show that frustrated magnetism emerges at a moderate correlation strength of $U\approx 3$ eV. Our classical Monte-Carlo simulations in the experimentally extracted parameter range show that the system is susceptible to forming long-period magnetic structures.  This may allow, for example, the Bi--Te layers containing topological fermions to be subjected to a variety of helimagnetic or topological skyrmionic structures \cite{Leonov15} under suitable perturbations, such as chemical substitution. 
\begin{figure*}
\includegraphics[width=1.0\linewidth]{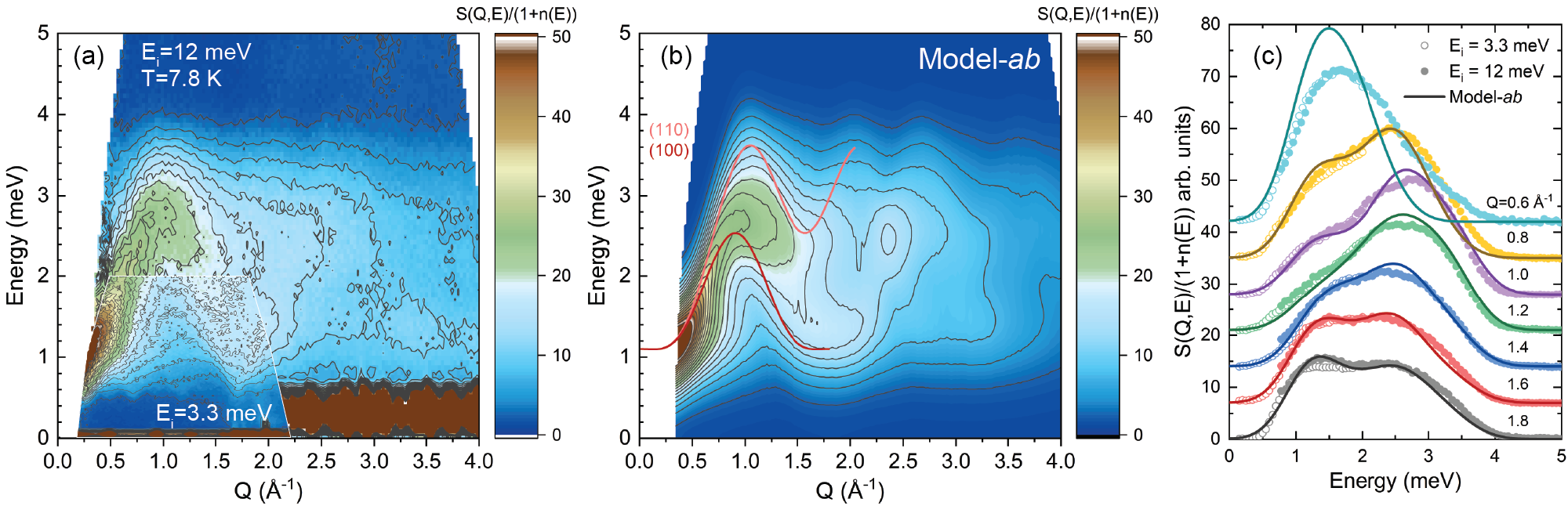}
\caption{\footnotesize (a) Inelastic neutron scattering intensities of both the $E_{i}=$ 12 meV and $E_{i}=$ 3.3 meV data plotted versus $Q$ and $E$ from a powder sample of MnBi$_{2}$Te$_{4}$ in the ordered AF phase at $T=$ 7.8 K. (b) Results of Heisenberg model calculations of the intralayer excitations (Model-$ab$) where pink and red lines are the dispersion in the (110) and (100) directions, respectively.  (c) Cuts of the neutron intensity at different $Q$-values for $E_{i}=$ 12 meV (solid circles) and 3.3 meV (empty circles) as compared to  Model-$ab$ calculations (lines).}
\label{fig1}
\end{figure*}

INS measurements on powder samples of MnBi$_{2}$Te$_{4}$ were performed on the Cold Neutron Chopper Spectrometer (CNCS) at the Spallation Neutron Source at Oak Ridge National Laboratory.  The powder sample of MnBi$_{2}$Te$_{4}$ used for this study was synthesized by annealing at 585 C for a week the homogeneous stoichiometric mixture of the elements quenched from 900 C. Magnetic measurements and powder neutron diffraction confirm long-range $A$-type AF order below $T_{N}=$ 24 K \cite{Yan19}.  Eight grams of powder sample were loaded into 1/2" diameter aluminum can and attached to a closed-cycle refrigerator for measurements below and above $T_{N}$ at $T=$ 7.8 K and 30 K, respectively, using incident neutron energies of $E_{i}=$ 3.3 and 12 meV.  The intensities are plotted as $S(Q,E)/(1+n(E))$ where $Q$ is the momentum transfer, $E$ is the energy transfer,  and $n(E)=(\mathrm{exp}(E/k_{B}T)-1)^{-1}$ is the Bose population factor. This intensity is proportional to the imaginary part of the dynamical susceptibility times the square of the magnetic form factor, $f^{2}(Q)\chi''(Q,E)$.  Other data treatment details are described in the Supplementary Material ($SM$) \cite{SM}.

Figure 1(a) shows the $Q$ and $E$ dependencies of the $E_{i}=$ 3.3 meV INS data superimposed on the 12 meV data measured in the AF phase at $T=$ 7.8 K.  The data show dispersing spin wave excitations that emanate from $Q \approx 0$, reach a maximum energy of $E \approx$ 3.5 meV near the Brillouin zone boundary of the triangular layer at $Q \approx$ 1 \AA$^{-1}$, and return to a finite energy due to a spin gap near the magnetic/nuclear (1,0,$L$) zone centers at $Q \approx$ 1.7 \AA$^{-1}$.  The signal weakens for larger $Q$ due to the magnetic form factor.  The magnetic spectral features are very broad and the 3.3 meV and 12 meV data sets are nearly indistinguishable despite the sizable difference in instrumental energy resolution [full-width-at-half-maximum (FWHM) of 0.15 meV and 0.7 meV, respectively].  This provides evidence that strong intrinsic sources of line broadening, such as magnon-phonon and/or magnon-electron coupling, are present in MnBi$_{2}$Te$_{4}$.  Surprisingly, the INS features are qualitatively similar to the FM-TI (Bi$_{0.95}$Mn$_{0.05}$)$_2$Te$_3$ \cite{Vaknin19} where dilute concentrations of Mn are expected to substitute randomly into Bi triangular layers. 
\begin{figure*}
\includegraphics[width=0.75\linewidth]{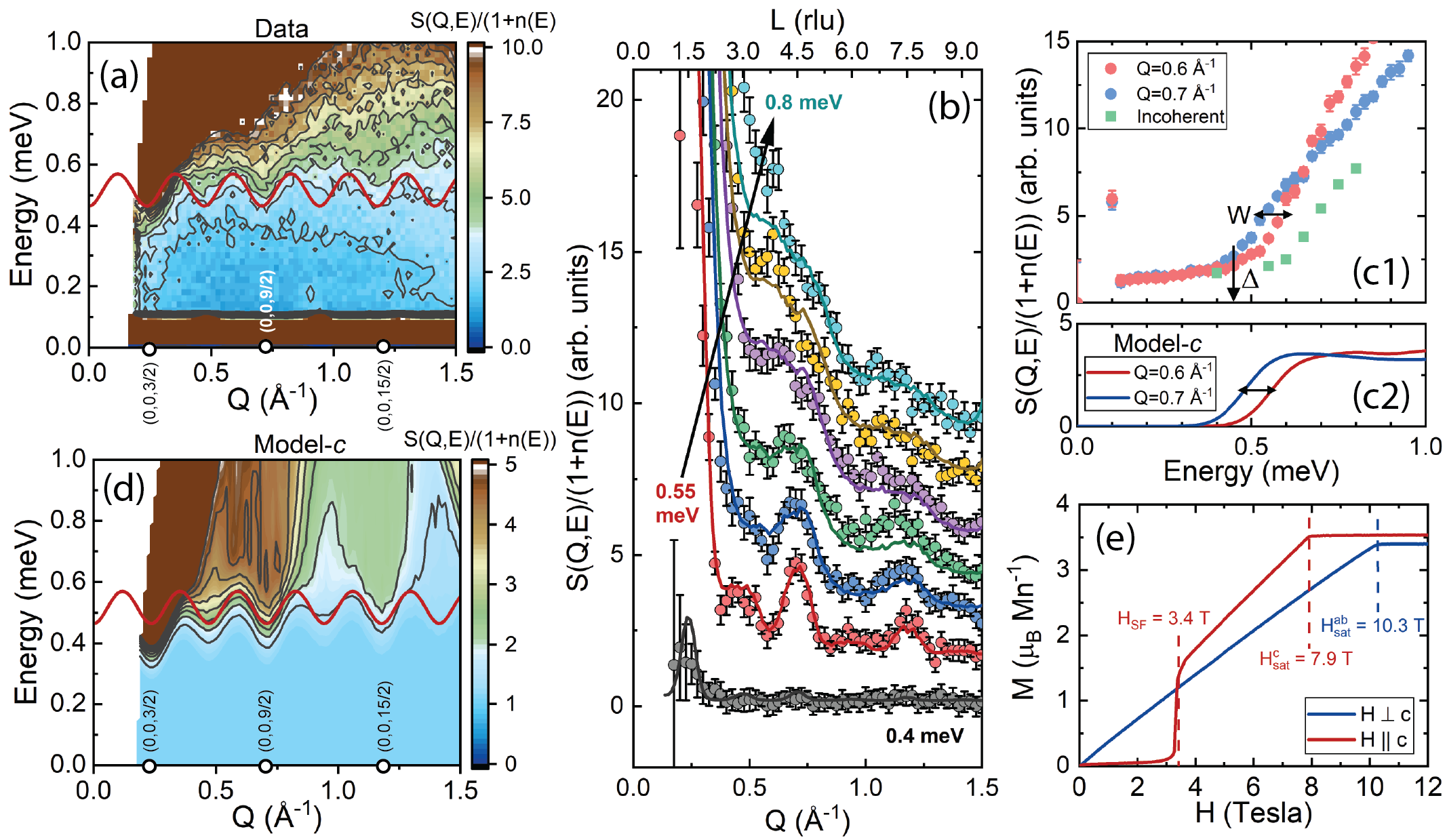}
\caption{\footnotesize (a) Inelastic neutron scattering intensities of MnBi$_{2}$Te$_{4}$ measured at $T=$ 7.8 K focused on the low energy gap edge with $E_{i}=$ 3.3 meV.  (b)  Several constant energy $Q$-cuts at the gap edge from the data (circles) and from Model-$c$ (lines).  Plots are vertically offset for clarity.  (c1) Low energy magnetic energy spectrum showing the spin gap ($\Delta$) near (0,0,9/2) ($Q=$ 0.7 \AA$^{-1}$, blue circles) and an estimate of the bandwidth ($W$) using a cut near the interlayer AF zone boundary ($Q=$ 0.6 \AA$^{-1}$, red circles).  Green squares are estimates of the incoherent background originating from intralayer spin wave modes. (c2) Same cuts as in (c1) obtained from Model-$c$.  (d) Numerical calculations of the INS intensity from Model-$c$. In (a) and (d), the red line shows the dispersion of spin wave modes along $c$ from Model-$c$. (e) Magnetization data from a single-crystal of MnBi$_{2}$Te$_{4}$ highlighting spin-flop and saturation fields.}
\label{fig2}
\end{figure*}

Figures 2(a-c) show the spin gap structure in more detail.  Despite the heavy broadening of the higher energy modes, Fig.~2(a) and constant energy $Q$-cuts in Fig.~2(b) find sharp dispersion minima at momenta of (0,0,$\frac{3}{2}$), (0,0,$\frac{9}{2}$), and (0,0,$\frac{15}{2}$), corresponding to the magnetic zone centers of the $A$-type AF structure. This observation suggests that interlayer interactions are not negligible, which is surprising given the large spacing of 13.6 \AA\ between Mn layers.  An energy cut at the dispersion minimum at (0,0,$\frac{9}{2}$) ($Q=$ 0.7 \AA$^{-1}$) in Fig.~2(c1) indicates a spin gap with an onset of $\Delta \approx$ 0.5 meV consistent with sizable uniaxial magnetic anisotropy.

The quantitative details of the magnetic interactions become more apparent based on fitting the data to a local-moment Heisenberg model,
\begin{equation}
H= -J_{1} \sum_{\langle ij \rangle||} \textbf{S}_{i} \cdot \textbf{S}_{j} - J_{2} \sum_{\langle\langle ij \rangle\rangle||} \textbf{S}_{i} \cdot \textbf{S}_{j} - J_{c} \sum_{\langle ij \rangle\perp}\textbf{S}_{i} \cdot \textbf{S}_{j} - D \sum_{i} S_{i,z}^{2}
\label{heisenberg}
\end{equation}
where $J_{1}$ and $J_{2}$ correspond to nearest-neighbor (NN) and next-nearest-neighbor (NNN) interactions within a single triangular layer, $J_{c}$ corresponds to an AF NN interlayer coupling, and $D>0$ is the uniaxial anisotropy.  Here $J_{i}>0$ corresponds to FM coupling.  As we describe below, the sharpness of intralayer modes (at the gap edge) and the broad, high energy interlayer modes cannot be consistently modeled with a single set of Heisenberg parameters.  Therefore, we develop independent models to describe the gap edge (Model-$c$) and high energy data (Model-$ab$).

Model-$c$ captures the interlayer interaction $J_{c}$ and uniaxial anisotropy $D$ from our magnetization and gap edge INS data.  The magnetization was measured at $T =$ 2 K on single-crystal specimens, shown in Fig.~2(e), and reveal spin-flop and saturation fields $H_{SF}=$ 3.4 T and $H_{sat}^{c}=$ 7.9 T  with $H || c$ and $H_{sat}^{ab}=$ 10.3 T with $H || ab$, consistent with previous reports \cite{Otrokov18_2,Lee18,Yan19}. Within the Heisenberg model and starting from $A$-type order with moments along $c$, these critical fields are given by the expressions $g\mu_{B}H_{SF}=2S \sqrt{D(6|J_{c}|-D)}$, $g\mu_{B}H_{sat}^{c}=2S(6|J_{c}|-D)$, and $g\mu_{B}H_{sat}^{ab}=2S(6|J_{c}|+D)$ (where $g\approx2$ and $S \approx 5/2$) and provide a range of values for $SD \approx$ 0.07 -- 0.1 meV and $-SJ_{c} \approx$ 0.08 -- 0.09 meV.   The magnetization data provide an estimate for $\Delta = 2S\sqrt{D(6|J_{c}|+D)}=$ 0.4 -- 0.5 meV that is consistent with the INS data in Fig.~2(c1).  

We also analyze $SD$ and $SJ_{c}$ by comparing the gap edge INS data to calculations of the powder-averaged spin wave intensities following the procedure outlined in Ref.~ \cite{McQueeney08}. We assume resolution-limited features ($FWHM=$ 0.15 meV) and fix $SJ_{1}$ and $SJ_{2}$ to nominal values since the energies are too low to effectively fit these parameters.  We then vary $SJ_{c}$ and $SD$ and compare the calculated spin wave intensities to a series of constant-energy $Q$-cuts from 0.4 -- 0.8 meV, as shown in Fig.~2(b).  Much better agreement with the data is obtained by the addition of incoherent background contributions that presumably originate from the broad, intralayer excitations described below. The resulting $\chi^2$ goodness-of-fit displays a rather shallow minimum that does not allow precise determination of $SD$ and $SJ_{c}$ (see Fig.~S4 in $SM$ \cite{SM}) and deviates somewhat from the values determined from the magnetization data.

Within the shallow minimum in $\chi^2$, a representative set of parameters can be ascertained from the spin gap and the bandwidth of interlayer excitations ($W$) shown in Fig.~2(c1).  The bandwidth is determined by the energy at the AF zone boundary at $Q=$ 0.6 \AA$^{-1}$ where $W=6S|J_{c}|+2SD-\Delta \approx$ 0.1 meV.  $\Delta$ and $W$ provide rough estimates of $SD \approx$ 0.12 meV and $SJ_{c} \approx$ -0.055 meV that sit within the minimum in $\chi^2$ and Figs.~2(a)-(d) shows these parameters provide a good representation of the gap edge data.  Full details of the Model-$c$ fits can be found in the $SM$ \cite{SM}.

We now turn to the development of Model-$ab$ for the intralayer spin dynamics.  Optimal values of $SJ_{1}$, $SJ_{2}$, $SD$ were determined by sampling these parameters over a regular mesh and comparing to the full magnetic spectrum of the combined $E_{i}=$ 3.3 and 12 meV data summed over the momentum range from $Q=0.8 - 1.9$ \AA$^{-1}$.  Best fits are obtained when calculated spectra are convoluted in energy by 0.85 meV, which is larger than the instrumental elastic resolution and points to significant lifetime broadening of the intralayer spin waves. $SJ_{c}$ has little effect on the full magnetic spectrum and was set to zero.  Details of the Model-$ab$ fits can be found in the $SM$ \cite{SM}.

The Model-$ab$ fit converges to an optimal value of $SD \approx 0.55 $ meV, corresponding to an effective spin gap of $\Delta =$ 1.1 meV for the intralayer excitations.  This value differs from the true spin gap of $\approx$ 0.5 meV obtained from magnetization and gap edge data.  This source of this discrepancy is unknown, but may arise from strongly \textbf{Q}-dependent lifetime broadening that cannot be extracted from powder data.  The optimal intralayer exchange energies are $SJ_{1}=0.28(2)$ meV and $SJ_{2}=-0.10(2)$ meV. The value of $SJ_{1}$ for MnBi$_{2}$Te$_{4}$ is consistent with that obtained from single-crystal INS studies of hexagonal MnTe \cite{Hennion06}, which has similarly stacked Te-Mn-Te triangular layers.  The large AF $J_{2}$ is a frustrating interaction that moves magnetic spectral weight from the top of the band down to lower energies [see Fig.~S6 \cite{SM}].  Model-$ab$ provides an excellent representation of the intralayer excitations, as demonstrated by comparisons to the data in Figs.~1(a)--(c).  Agreement of Model-$ab$ with the data is less satisfactory at low-$Q$ where the data display a high energy tail [see Fig.~1(c)].  This disagreement may be caused by low-angle background and linewidth effects that go beyond our simple assumption of a uniform broadening parameter.  Table I summarizes the parameters of Model-$c$ and Model-$ab$.

\begin{table}
\caption {Heisenberg model parameters obtained from magnetization data and low energy INS data (Model-$c$), high energy INS data (Model-$ab$), and DFT+$U$ calculations. The broadening parameter for INS data (FWHM) is also provided.  All values are in meV.}
\renewcommand\arraystretch{1.25}
\begin{ruledtabular}
\begin{tabular}{ c | c | c | c | c | c }
~  ~   &  ~$SJ_{1}$ ~ &  ~$SJ_{2}$~ & ~$SJ_{c}$~  &~$SD$~  & ~FWHM~  \\
\hline
Model-$c$ (Mag.)			& -- & -- & -0.085(5) & 0.085(15) & --  \\
Model-$c$	 (INS)			&  0.23 & 0 & -0.055 & 0.12 & 0.15  \\
Model-$ab$ 			& 0.28(2) & -0.1(2) & 0 & 0.55(5)  & 0.85  \\
DFT+U (5 eV)			&  0.7 & -0.05 & -0.025 & 0.13 & --   \\
DFT+U (3 eV)			&  0.55 & -0.14 & -0.087 & 0.18 &  --  \\
\end{tabular}
\end{ruledtabular}
\label{CW_table}
\end{table}

Our major finding is that competing interactions within the triangular layer are significant ($|J_{2}/J_{1}| = 0.36$) and close to the classical instability limit for intralayer FM ground state ($|J_{2}/J_{1}| = 1/3$).  While the system resides on the FM side, chemical doping can possibly induce helical or skyrmion phases that are expected nearby. To quantify this expectation, we have calculated the magnetic phase diagram (including a magnetic field) using classical Monte-Carlo (MC) simulations.  In Fig.~3(a), we show the low-$T$ phase diagram for the experimentally found anisotropy value $D/J_{1} = 0.4$ as a function of $|J_2/J_1|$ and magnetic field $h/J_1$ along the $z$-direction. Vertical spiral, skyrmion and up-up-down-down stripe phases [see Figs.~3(b-d)] appear at a slightly larger frustration ratio of $|J_2/J_1| \geq 0.5$ than our INS data suggests.  MC simulations also find skyrmion phases at appear at smaller anisotropy values $D/J_{1} \lesssim 0.1$ at $|J_{2}/J_{1}| = $ 0.4 (see $SM$ \cite{SM}).  This raises the possibility for spiral or skyrmion phases to appear, for example, in Sb-substituted Mn(Bi,Sb)$_2$Te$_4$ where $D$ is found to be significantly smaller \cite{Yan19_2}. 

\begin{figure}
\includegraphics[width=0.9\linewidth]{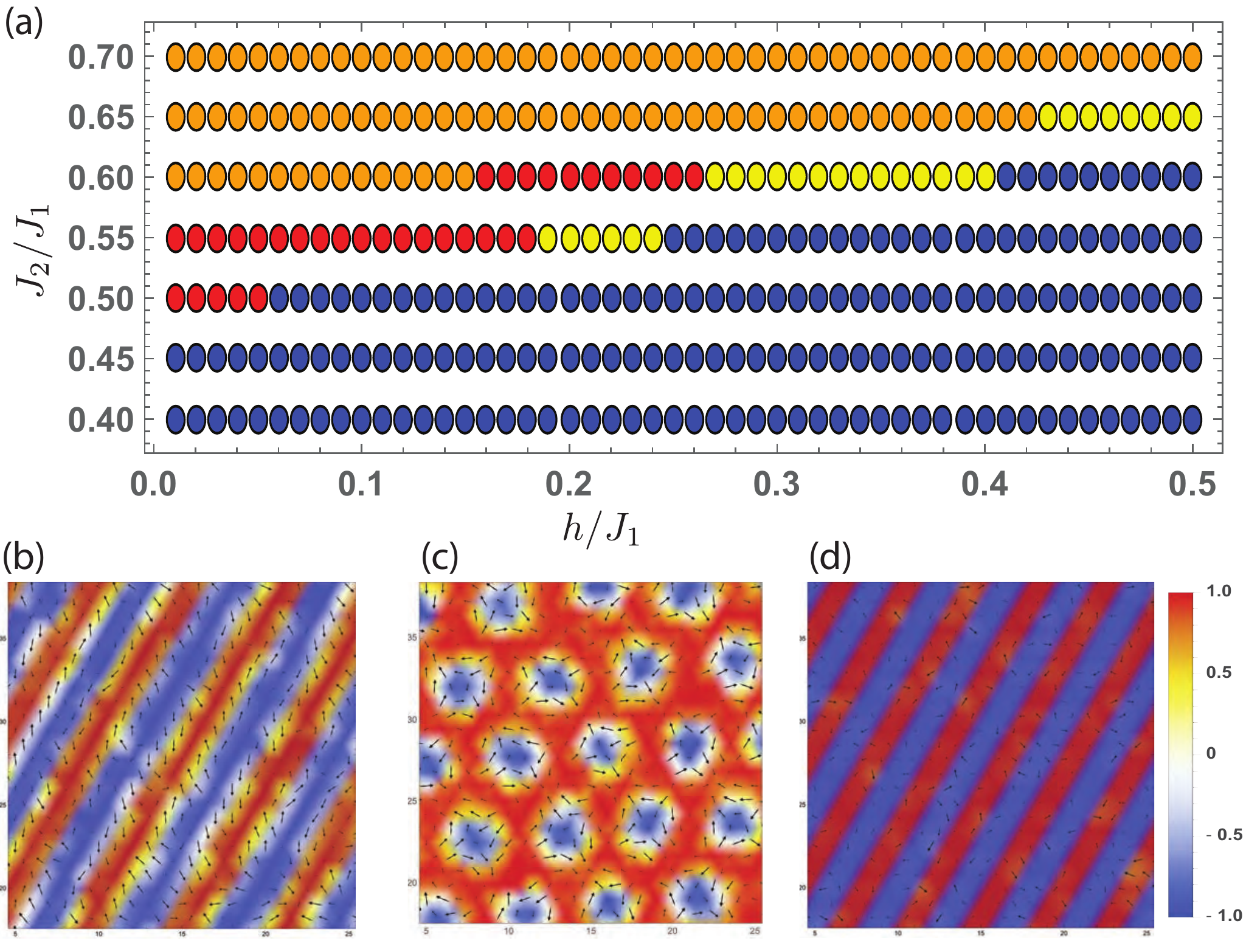}
\caption{\footnotesize (a) Low-temperature magnetic phase diagram as a function of $|J_2/J_1|$ and magnetic field $h/J_1$ for fixed anisotropy $D/J_1 = 0.4$ and temperature $T = 0.08 J_1$. Different phases are polarized paramagnet (blue), vertical spiral (red), multi-$q$ (skyrmion) phase (yellow), and up-up-down-down (orange). (b-d) Real-space spin configurations of vertical spiral (b), multi-$q$ (skyrmion) crystal (c), and up-up-down-down phases (d). Color denotes $S^z$ component (scale bar shown) and arrows denote the in-plane components $(S^x, S^y)$.}
\label{fig3}
\end{figure}

Recent first-principles electronic structure calculations with $U=$ 5 eV predict that $|J_{2}/J_{1}|<0.03$ \cite{Otrokov18_2,Otrokov19} in contradiction to our INS data. Here, we extract the Heisenberg parameters by performing an analysis of the energies of four ordered spin states \cite{Xiang13} based on DFT+$U$ calculations \cite{Dudarev98}.  Fig.~4(a) reveals that the value of $U$ is critical.  For example, $U \geq $ 2 eV is required to obtain the $A$-type ground state of MnBi$_{2}$Te$_{4}$.  While the DFT+$U$ values for the exchange are generally larger than the experimental values, Fig.~4 and Table \ref{CW_table} show that the ratios of $J_{2}/J_{1}$, and $J_{c}/J_{1}$ at moderate values of $U \approx$ 2.7 eV are consistent with the INS data. Further details of the computational methods in full can be found in the $SM$ \cite{SM}.  
\begin{figure}
\includegraphics[width=0.6\linewidth]{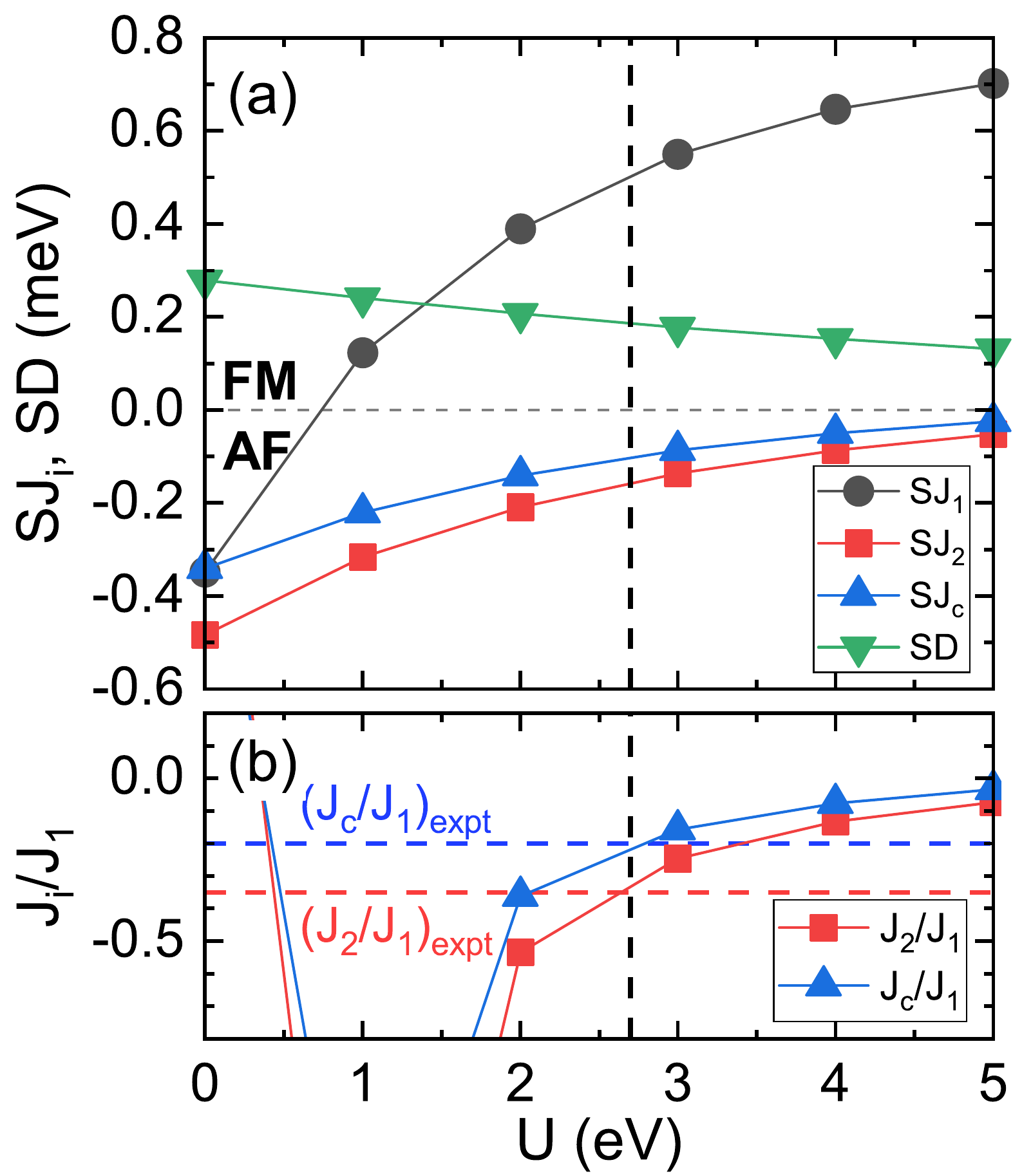}
\caption{\footnotesize First-principles calculations of (a) Heisenberg parameters and (b) ratios of key exchange interactions versus $U$ using PBE. In (b), the red and blue horizontal dashed lines correspond to experimental values for $J_{2}/J_{1}$ and $J_{c}/J_{1}$ and the vertical black dashed line shows the best value of $U \approx 2.7$ eV.}
\label{fig4}
\end{figure}

Overall, our findings indicate that AFTI MnBi$_{2}$Te$_{4}$ shows elements of frustration ($|J_{2}/J_{1}| \approx 1/3$), Ising anisotropy (taking $SD$ from Model-$c$ and $SJ_{1}$ from Model-$ab$ suggests 0.3 $< D/J_{1} <$ 0.4), and metamagnetism ($|J_{c}|/D > 0.5$).  The presence of low-field metamagnetism in MnBi$_{2}$Te$_{4}$ is similar to that found in $MX_{2}$ transition metal halide triangular lattice antiferromagnets \cite{McGuire17}. Componds such as FeCl$_{2}$ \cite{Birgeneau72} and FeBr$_{2}$ \cite{Yelon75}, also display strongly competing $J_{1}-J_{2}$ interactions within the triangular layer and can host multi-$q$ magnetic structures \cite{McGuire17}. $MX_{2}$ compounds have therefore been proposed to host skyrmion phases in applied fields \cite{Leonov15}.  In MnBi$_{2}$Te$_{4}$, similar frustration could lead to skyrmion phases and may also result in complex spin textures near the sample surface where magnetic interactions may be modified by strain or surface termination effects.  This could explain recent ARPES \cite{Hao19,Swatek19} and thin film magnetization data \cite{Gong18}  that are not consistent with uniformly FM layers.  

Even from our powder samples, we find clear evidence for strongly \textbf{Q}-dependent broadening, which should be investigated in INS studies of single-crystal samples.  Such lifetime broadening could be related to frustration or to coupling between magnetic fluctuations and charge carriers, as inferred from magnetotransport measurements \cite{Yan19_2}.

\section{Acknowledgments}
Authors would like to thank Q.~Zhang for powder diffraction measurements.  Work at Oak Ridge National Laboratory (ORNL) and Ames Laboratory was supported by the U.S. Department of Energy (USDOE), Office of Basic Energy Sciences, Division of Materials Sciences and Engineering. A portion of this research used resources at the Spallation Neutron Source, a USDOE Office of Science User Facility operated by ORNL.  Ames Laboratory is operated for the USDOE by Iowa State University under Contract No. DE-AC02-07CH11358.


\begin{thebibliography}{99}
\bibitem{Tokura19} Y. Tokura, K. Yasuda, and A. Tsukazaki, Nat. Rev. Phys. {\bf 1}, 126 (2019).
\bibitem{Zhang13} J. Zhang, C.-Z. Chang, P. Tang, Z. Zhang, X. Feng, K. Li, L.-l. Wang, X. Chen, C. Liu, W. Duan \textit{et al.} Science \textbf{339}, 1582 (2013).
\bibitem{Chang13} C.-Z. Chang, J. Zhang, X. Feng, J. Shen, Z. Zhang, M. Guo, K. Li, Y. Ou, P. Wei, L.-L. Wang \textit{et al.} Science \textbf{340}, 167 (2013).
\bibitem{Chang15} C.-Z. Chang, W. Zhao, D. Y. Kim, H. Zhang, B. A. Assaf, D. Heiman, S.-C. Zhang, C. Liu, M. H. W. Chan, and J. S. Moodera, Nat. Mater. \textbf{14}, 473 (2015).
\bibitem{Otrokov18} M. M. Otrokov, T. V. Menshchikova, M. G. Vergniory, I. P. Rusinov, A. Y. Vyazovskaya, M. K. Yu, G. Bihlmayer, A. Ernst, P. M. Echenique, A. Arnau \textit{et al.}, 2D Materials \textbf{4}, 025082 (2017).
\bibitem{Zhang18} D. Zhang, M. Shi, K. He, D. Xing, H. Zhang, and J. Wang, arXiv:1808.08014 (2018).
\bibitem{Otrokov18_2} M. M. Otrokov, I. I. Klimovskikh, H. Bentmann, A. Zeugner, Z. S. Aliev, S. Gass, A. U. B. Wolter, A. V. Koroleva, D. Estyunin, A. M. Shikin \textit{et al.}, arXiv:1809.07389 (2018).
\bibitem{Lee18} S.-H. Lee, Y. Zhu, Y. Wang, L. Miao, T. Pillsbury, S. Kempinger, D. Graf, N. Alem, C. Chang, N. Samarth, and Z. Mao, arXiv:1812.00339 (2018).
\bibitem{Otrokov19} M. M. Otrokov, I. P. Rusinov, M. Blanco-Rey, M. Hoffmann, A. Y. Vyazovskaya, S. V. Eremeev, A. Ernst, P. M. Echenique, A. Arnau, and E. V. Chulkov, Phys. Rev. Lett. \textbf{122}, 107202 (2019).
\bibitem{Yan19} J.-Q. Yan, Q. Zhang, T. Heitmann, Z. Huang, K. Y. Chen, J. G. Cheng, W. Wu, D. Vaknin, B. C. Sales, and R. J. McQueeney Phys. Rev. Mater. \textbf{3}, 064202 (2019).
\bibitem{Moore10} R. S. K. Mong, A. M. Essin, and J. E. Moore, Phys. Rev. B \textbf{81}, 245209 (2010).
\bibitem{Gong18} Y. Gong, J. Guo, J. Li, K. Zhu, M. Liao, X. Liu, Q. Zhang, L. Gu, L. Tang, X. Feng \textit{et al.}, Chin. Phys. Lett. \textbf{36}, 076801 (2019).
\bibitem{Zhang19} Y. Deng, Y. Yu, M.-Z. Shi, J. Wang, X.-H. Chen, and Y. Zhang, arXiv:1904.11468 (2019).
\bibitem{Tanaka76} Y. Tanaka and N. Uryu, Progr. Theor. Phys. \textbf{55}, 1356 (1976).
\bibitem{Murao96} K. Murao, F. Matsubara, and T. Kudo, J. Phys. Soc. Jpn. \textbf{65}, 1399 (1996).
\bibitem{Leonov15} A. O. Leonov and M. Mostovoy, Nat. Commun. \textbf{6}, 8275 (2015).
\bibitem{SM} Supplementary Material reference.
\bibitem{Vaknin19} D. Vaknin, D. M. Pajerowski, D. L. Schlagel, K. W. Dennis, and R. J. McQueeney, Phys. Rev. B \textbf{99}, 220404 (2019).
\bibitem{McQueeney08} R. J. McQueeney, J. Q. Yan, S. Chang, and J. Ma, Phys. Rev. B \textbf{78}, 184417 (2008).
\bibitem{Hennion06} W. Szuszkiewicz, E. Dynowska, B. Witkowska, and B. Hennion, Phys. Rev. B \textbf{73}, 104403 (2006).
\bibitem{Yan19_2} J.-Q. Yan, S. Okamoto, M. A. McGuire, A. F. May, R. J. McQueeney, and B. C. Sales \textit{et al.}, arXiv:1905.00400 (2019).
\bibitem{Xiang13} H. Xiang, C. Lee, H.-J. Koo, X. Gong, and M.-H. Whangbo, Dalton Trans. \textbf{42}, 823 (2013).
\bibitem{Dudarev98} S. L. Dudarev, G. A. Botton, S. Y. Savrasov, C. J. Humphreys, and A. P. Sutton, Phys. Rev. B. \textbf{57}, 1505 (1998).
\bibitem{McGuire17} A. M. McGuire, Crystals \textbf{7}, 121 (2017).
\bibitem{Birgeneau72} R. J. Birgeneau, W. B. Yelon, E. Cohen, and J. Makovsky, Phys. Rev. B \textbf{5}, 2607 (1972).
\bibitem{Yelon75} W. B. Yelon, Oelete, and C. Vettier. J.Phys. C: Solid State Phys. \textbf{8}, 2760 (1975).
\bibitem{Hao19} Y.-J.Hao, P. Liu, Y. Feng, X.-M. Ma, E. F. Schwier, M. Arita, S. Kumar, C. Hu, R. Lu, M. Zeng \textit{et al.}, arXiv:1907.03722 (2019).
\bibitem{Swatek19} P. Swatek, Y. Wu, L. L. Wang, K. Lee, B. Schrunk, J.-Q. Yan, and A. Kaminski, arXiv:1907.09596 (2019).

 

\end{thebibliography}
\end{document}